\documentclass[useAMS,usenatbib]{mn2e}
\newcommand{\etavalue}{$\eta=11\%$}
\newcommand{\gammavalue}{$\Gamma_0=2\times10^3$}
\newcommand{\deltagammavalue}{$\Delta\Gamma=2\times10^3$}
\usepackage{natbib}
\usepackage{amsmath}
\usepackage{amssymb}
\usepackage{graphicx}
\usepackage{color}
\usepackage{multirow}
\usepackage{color}
\usepackage{enumerate}
\usepackage{placeins}
\usepackage[table]{xcolor}
\usepackage{float}
\usepackage{comment}
\usepackage{enumerate}
\usepackage{adjustbox}
\usepackage{subfigure}
\begin{document}
\title{Modeling the GeV emission of HESS J0632+057}
\author[S. X. Yi et al.]{ Shu-Xu Yi$^{\dagger}$ , K.S. Cheng$^{\dagger}$ \\
$^{\dagger}$ Department of Physics, the University of Hong Kong University, Pokfulam Road, Hong Kong\\
}
\date{\today}
\maketitle
\begin{abstract}
The binary system HESS J0632+057 was recently detected by {Fermi} to possess orbital modulated GeV emission. In this paper, we study the possibility that the compact companion of HESS J0632+057 is a pulsar. Under such a presumption, we focus on the high energy emission mechanism of this system, which is as follows. The pulsar companion travels through the circumstellar disc of the main sequence star twice in each orbit, when some of the matter is gravity-captured. The captured matter develops an accretion disc around the pulsar, and the soft photons from which are inverse Compton scattered by the pulsar wind as the GeV emission from the system. With proper choice of parameters, SED and light curve which are in accordance with observations can be produced. We predict that the light curve of GeV emission has two peaks, the larger one is at around 0.4 after the periastron (or 0.1 after the X-ray maximum), while the smaller one is between phases 0 and 0.1, with integrated flux one forth of the larger one.
\end{abstract}

\section{Introduction}
HESS J0632+057 was discovered as a TeV gamma-ray source \citep{2007A&A...469L...1A}. The follow-up observations found periodic time variations of emission in X-ray and TeV  \citep{2007A&A...469L...1A,2009ApJ...690L.101H,2009ApJ...698L..94A,2012ApJ...754L..10A,2014ApJ...780..168A}, which revealed the binary nature of this system. This system is considered to be one of the $\gamma$-ray binary systems\footnote{The term ``$\gamma$-ray binary" in our paper is equivalent to ``$\gamma$-ray emitting binary" with a pulsar or a black hole companion in \citealt{2015CRPhy..16..661D}. While the author use ``$\gamma$-ray binary" to specifically refer the case when the compact companion is a pulsar.}, (see \citealt{2015CRPhy..16..661D} for a review). The optical companion is a Be star MWC 148 \citep{2007A&A...469L...1A}. \cite{2012MNRAS.421.1103C} detected the radial velocity variations in the photospheric lines of the optical companion and thus determined the period of the binary system $P=321\pm5$\,days and the mass function $f(M)=0.35^{+0.20}_{-0.15}\,M_\odot$. \cite{2014ApJ...780..168A} further refined the orbital period to be $315^{+6}_{-4}$\,days. The mass of the optical companion was determined to be $13.2-19.0\,M_\odot$ \citep{2010ApJ...724..306A}. Due to the uncertainty of the inclination of the orbit, the possible mass range of the compact companion is determined to lie in the range $1.3-7.1\,M_\odot$, which covers two distinct categories: neutron stars (usually believed to have mass $M_c<3\,M_\odot$, \citealt{1996ApJ...470L..61K}) or black holes. We summary other aspects of the binary in the section II.

There are two families of mechanisms to explain the high energy emission of $\gamma$-ray binary systems, depending on the nature of the compact companion. When the compact object is a black hole, accretion from the the main sequence star is expected to produce a microquasar jet, which accounts for the high energy emission (see a review by \citealt{2009IJMPD..18..347B} and references therein). On the other hand, if the compact object is a neutron star (a pulsar), one main branch of theories explain the high energy emissions to originate from the collision shock between the pulsar wind and the stellar wind/circumstellar disc (CD) \citep{1981MNRAS.194P...1M,2006A&A...456..801D}.

An example for the pulsar wind scenario is in the $\gamma$-ray binary system PSR B1259-63/LS2883, where the compact companion is known to be a pulsar for its radio pulsations. Theoretical models on this system (\citealt{2011MNRAS.416.1067K} and references therein) proposed that the X-ray are produced by synchrotron radiation from shock-heated electrons, and TeV emission is produced via Inverse Compton (IC) process of the hot electrons (both synchrotron self-Compton and external IC scattering the thermal photons from the optical companion). However to explain the GeV emission (HE, 100\,MeV-100\,GeV) is more difficult. Many of the efforts try to explain the GeV emission with IC scattering by cold electron from the pulsar wind (e.g., \citealt{2012ApJ...752L..17K,2013A&A...557A.127D}). The main difficulty is the lackness of soft photons, because the GeV peaks at $\sim20$\,days after periastron, where neither the circumstellar matter nor the soft photon field from the optical companion is at the densest. Recently we use a new approach to the GeV emission of the system  (cf. \citealt{2017ApJ...836..241T,Yi2017}). In that framework, the soft photons are from the accretion disc around the pulsar, which was developed after the CD matter being gravity-captured by the pulsar at disc-crossing. The time required for the accretion disc to form corresponds to the lag-time between the X-ray/TeV maximum and the GeV peak.

HESS 0632+057 seems to share some observational similarities, therefore people put efforts in searching for its GeV emission for a long time. With 3.5 years of data from {\it Fermi} Large Area Telescope (LAT), \cite{2013MNRAS.436..740C} found no GeV emission and set the upper limit of the GeV flux from this system. A recent search for HESS J0632+057 with up-to-date {\it Fermi}-LAT data discovered the orbital-modulated HE GeV emission \citep{Li2017}. With the studies on the Balmer and Fe II line profile variations from the CD, \cite{2015ApJ...804L..32M} concluded that the pulsar wind mechanism is the most likely candidate for high energy emissions in this system. These motivate us to apply the model proposed for B1259-63/LS2883 to HESS J0632+057.

 The presupposition of the model is that the compact object in HESS J0632+057 has a mass of a neutron star. In the section II, we discuss on the mass of the compact object, based on the present observational constraints. We also discuss other parameters of the binary system which will be used in the next section. In the section III we discuss the mass transfer and accretion process. In section IV, we calculated the time-dependent HE SED and light curve, as a prediction of our model for future test by observations. We conclude our finding and discuss in the final section.
\section{MODEL PARAMETERS}
In our model, the compact companion of the binary system is a pulsar. When the pulsar travels through the CD, some of the materials are gravity-captured by it. An accretion disc then develops around the neutron star, and the soft photons from the accretion disc are IC scattered by the relativistic electrons in the pulsar wind. The up-scattered photons are the observed HE emission.

\subsection{The parameters of the binary and the circumstellar disc}
To explore the validity of the proposed scenario, we first examine the parameters of the system within ranges set by previous observations. Those parameters include the masses of the compact companion $M_{\rm{c}}$ and optical companion $M_\star$, the configration of the orbit, the geometry and the density profile of the CD, the spin property of the pulsar and the velocity distribution of the pulsar wind.

We start with ranges of the parameters of the optical companion and the orbital solution of \citep{2010ApJ...724..306A,2012MNRAS.421.1103C}: $K=22.5\pm5.7$\,km/s, $M_\star=16.1\pm2.9\,M_\odot$, $e=0.83\pm0.08$, $P=321\pm5$\,days, where $K$ is the amplitude of the radial velocity modulation, $e$ and $P$ are the eccentricity and period of the orbit. The resulted mass function $f(M)=0.35^{+0.20}_{-0.15}\,M_\odot$, where:
\begin{equation}
f(M)\equiv\frac{PK^3}{2\pi G}(1-e^2)^{3/2}=M_{\rm{c}}^3\sin^3i_{\rm{o}}/(M_\star+M_{\rm{c}})^2.
\end{equation}
In the above equation, $i_{\rm{o}}$ is the inclination of the orbit. \cite{2012MNRAS.421.1103C} limited $i_{\rm{o}}$ to be within $47^\circ\sim80^\circ$.

\begin{figure}
\centering
\includegraphics[width=7cm]{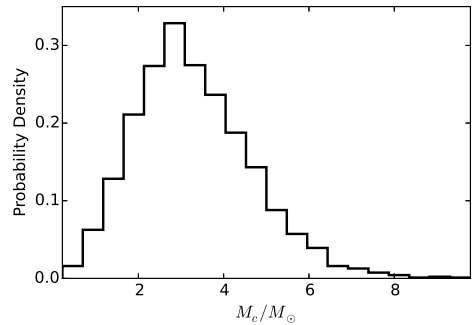}
\caption{The probability density distribution of $M_{\rm{c}}$, based on the observed distribution of $K$, $e$, $P$ and $M_\star$.}\label{fig:1}
\end{figure}

Suppose that the distribution of $K$, $M_\star$ and $e$ are Gaussian and the standard derivative are their given uncertainty (with $P$ fixed to 321 days as in \citealt{2012MNRAS.421.1103C}), and $\sin i_{\rm{o}}$ is uniformly distributed in its range, we can obtain the distribution of $M_{\rm{c}}$. The probability density distribution of $M_{\rm{c}}$ is plotted in figure \ref{fig:1}. We can see from this figure that, the probability of the compact companion to have a mass of a neutron star ($<3\,M_\odot$) is less (than that of a black hole) but still significant.

We specify three sets of ($K$, $e$, $P$, $i_{\rm{o}}$, $M_\star$) in Table \ref{tab:para}. Under each parameters $M_{\rm{c}}=1.5\,M_\odot$, $2.0\,M_\odot$ and $2.4\,M_\odot$ respectively. The possibilities of $M_{\rm{c}}\le1.5\,M_\odot$, $2.0\,M_\odot$ and $2.4\,M_\odot$ are $\sim$8\%, $17\%$ and $28\%$ respectively.

Next, we discuss the properties of the CD.
Outside the CD, we suppose the velocity of the stellar wind is radially and takes the form \citep{1988A&A...198..200W}:
\begin{equation}
v_{\rm{w}}(r)=10^8\,\text{cm/s}\times(1-R_\star/r)^{1.5},
\label{eqn:wdv}
\end{equation}
and the density profile is:
\begin{equation}
\rho_{\rm{w}}(r)=\frac{10^{-8}\,M_\odot\,\rm{yr}^{-1}}{4\pi r^2v_{\rm{w}}(r)};
\label{eqn:wdd}
\end{equation}

The density distribution of the CD has been modeled by previous researchers \citep{Lee1991,Porter1999,Carciofi2006} as:
\begin{eqnarray}
\rho_{\rm{CD}}&=&\rho_0\big(\frac{R_\star}{R}\big)^n\exp\big(\frac{-z^2}{2H^2}\big)\\ \nonumber
&=&\rho_0\big(\frac{R_\star}{R}\big)^n\exp\big(\frac{-(\phi-\phi_{\rm{CD}})^2}{2\Delta\phi^2}\big),
\label{eqn:rho}
\end{eqnarray}

where $n$ is in the range $3\sim3.5$, with 3.5 corresponds to a steady state isothermal outflow, $H$ is the scale height of the disc, $z$ is the coordinate in the vertical direction of the CD. The second part of above equation is the disc density profile projected to the orbital plane, where $\phi_{\rm{CD}}$ is the azimuthal angle at which the mid-plane of the CD intersects with the orbital plane, and $\Delta\phi$ is the half width of the CD projected on the orbital plane. \cite{2006MNRAS.367.1201C} proposed that $\phi_{\rm{cd},\pm}$ and $\Delta\phi$ corresponded to the peak position and the width of the X-ray and TeV light curves. We use $\phi_{\rm{cd},+}=170^\circ$ and $\Delta\phi=2.2^\circ$ based on the X-ray light curve from \cite{2011ApJ...737L..11B}. 
\begin{figure}
\centering
\includegraphics[width=8cm]{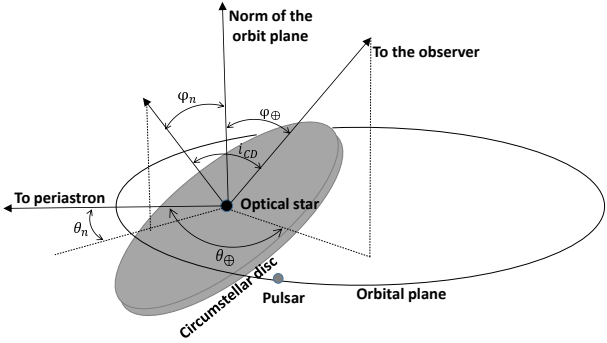}
\caption{Illustration of the binary system configuration. The $\varphi_{\rm{n}}$ and $\theta_{\rm{n}}$ illustrated in this figure do not represent to the values adopted in the text.}
\label{fig:sys}
\end{figure}
The orientation and inclination of the CD with respect to the orbit plane are important for the determination of the relative velocity between the neutron star and material on CD. Denoting the azimuth angle and the polar angle of the angular momentum vector of CD as $\varphi_{\rm{n}}$ and $\theta_{\rm{n}}$ respectively, and the azimuth angle and the polar angle of the direction to the observer as $\varphi_\oplus$ and $\theta_\oplus$, we have the following equation:
\begin{equation}
\sin\theta_\oplus\sin\theta_{\rm{n}}\cos(\varphi_\oplus-\varphi_{\rm{n}})+\cos\theta_\oplus\cos\theta_{\rm{n}}=\cos i_{\rm{CD}},
\label{eqn:earthpos}
\end{equation}
where $i_{\rm{CD}}$ is the inclination of the CD with respect to the observer (see figure \ref{fig:sys} for illustration). There is a relationship between $i_{\rm{CD}}$ and the full-width zero-intensity of certain absorption line of the inner region of the CD $\sigma_{\rm{FWZI}}$:
\begin{equation}
\sigma_{\rm{FWZI}}/2\times|\sin i_{\rm{CD}}|=(GM_\star/R_\star)^{1/2}.
\end{equation}
Using the measured $\sigma_{\rm{FWZI}}(\lambda5018)\sim1300$\,km/s and assuming the radius of MWC 148 to be $R_\star=6.6\,R_\odot$ (which is in the range of \citealt{2010ApJ...724..306A}), we obtain that $i_{\rm{CD}}=\pm73^\circ$. Take $i_{\rm{CD}}=\pm73^\circ$ and $\varphi_n=\phi_{\rm{CD},+}-90^\circ$ into equation (\ref{eqn:earthpos}), we have $\theta_{\rm{n}}=75.3^\circ$, $\varphi_{\rm{n}}=80^\circ$.
\subsection{The IC efficiency and the accretion rate of the disc}
We define the IC efficiency as the ratio between the luminosity of HE emission which originates from IC and the pulsar wind luminosity:
\begin{equation}
\xi=\frac{L_{\rm{IC}}}{L_{\rm{pw}}}.
\end{equation}
$L_{\rm{pw}}$ refers to the pulsar wind luminosity in particles, which could be much less than the spin-down power of pulsar near the pulsar if most spin-down power is still in electromagnetic wave \footnote{It is generally believed that the pulsar spin down power would be converted into pulsar wind power. However, at the IC location, how much pulsar spin down power has been converted into pulsar wind power is not known. As a result, the inferred pulsar wind power is a lower limit of the pulsar spin down power.}.
$\xi$ is determined by the density of the soft photons field, which is related with the accretion rate of the accretion disc.

The energy losing rate via IC of an electron through an isotropic photon field is:
\begin{equation}
\frac{dE_{\rm{e}}}{dt}=-\frac{4}{3}\sigma_{\rm{T}}cU_{\rm{ph}}\Gamma^2,\label{eqn:Edot}
\end{equation}
where $\sigma_{\rm{T}}$ is the Thompson scattering section, $U_{\rm{ph}}$ is the energy density of the photon field.
The above equation can be rewritten as the energy losing rate along the light of sight:
\begin{equation}
m_{\rm{e}}c^2\frac{d\Gamma}{dl}=-\frac{4}{3}\sigma_{\rm{T}}U_{\rm{ph}}\Gamma^2.\label{eqn:dE/dl}
\end{equation}
Integrating equation (\ref{eqn:dE/dl}), we have:
\begin{equation}
\frac{\Gamma_i}{\Gamma_f}=1+\frac{4\sigma_{\rm{T}}\Gamma_i}{3m_{\rm{e}}c^2}\int^\infty_0U_{\rm{ph}}(l)dl.
\label{eqn:ratio}
\end{equation}
The soft photons field are constituted by radiation from the accretion disc developed around the pulsar. In the axis of the accretion disc, the integration of $U_{\rm{ph}}$ along $l$ is:
\begin{equation}
\int^\infty_0U_{\rm{ph}}(l)dl=\frac{4\pi^5r_{\rm{in}}}{15c^3h^3}(kT_{\rm{in}})^4.
\label{eqn:integ}
\end{equation}
Take equation (\ref{eqn:integ}) into equation (\ref{eqn:ratio}),
\begin{equation}
\frac{\Gamma_i}{\Gamma_f}-1=7.43\times10^{-2}\frac{\Gamma_i}{10^3}\big(\frac{kT_{\rm{in}}}{10\,\text{eV}}\big)^4\frac{r_{\rm{in}}}{10^8\,\text{cm}}\approx\frac{L_{\rm{IC}}}{L_{\rm{pw}}-L_{\rm{IC}}}
\end{equation}
The last approximation relationship in the above equation is due to the fact that the energy of electrons are transferred to the photons via IC process, with an isotropic assumption.
If the accretion disc is an Shakura-Sunyaev disc \citep{1973A&A....24..337S}, then take the relationship between $r_{\rm{in}}$ and $T_{\rm{in}}$:
\begin{equation}
T=1.4\times10^4\alpha^{-\frac{1}{5}}\big(\frac{\dot{M}_{\rm{acc}}}{10^{16}\,\text{g/s}}\big)^{\frac{3}{10}}\big(\frac{M_{\rm{c}}}{M_\odot}\big)^{\frac{1}{4}}\big(\frac{r_{\rm{in}}}{10^{10}\,\text{cm}}\big)^{-\frac{3}{4}}\,\text{K},
\label{eqn:temper}
\end{equation}
then we have:
\begin{equation}
\frac{L_{\rm{IC}}}{L_{\rm{pw}}-L_{\rm{IC}}}=15.5\alpha^{-0.8}\frac{\Gamma_i}{10^3}\big(\frac{\dot{M}_{\rm{acc}}}{10^{16}\,\text{g/s}}\big)^{1.2}\frac{M_{\rm{c}}}{M_\odot}\big(\frac{r_{\rm{in}}}{10^8\,\text{cm}}\big)^{-2},
\label{eqn:ratio2}
\end{equation}
where $\alpha$ is the viscosity index in the SS disc.

If we take the Alfven radius $r_{\rm{A}}$ as $r_{\rm{in}}$,
\begin{equation}
\frac{r_{\rm{A}}}{10^8\,\text{cm}}=5.1\big(\frac{\dot{M}_{\rm{acc}}}{10^{16}\,\text{g/s}}\big)^{-2/7}\big(\frac{M_{\rm{c}}}{M_\odot}\big)^{-1/7}\mu_{30}^{4/7},
\label{eqn:alfven}
\end{equation}
where $\mu_{30}$ is the magnetic dipole in units of $10^{30}$\,G\,cm$^3$.
Therefore equation (\ref{eqn:ratio2}) becomes:
\begin{equation}
\frac{\xi}{1-\xi}=0.6\alpha^{-0.8}\frac{\Gamma_i}{10^3}\big(\frac{\dot{M}_{\rm{acc}}}{10^{16}\,\text{g/s}}\big)^{1.77}\big(\frac{M_{\rm{c}}}{M_\odot}\big)^{1.29}\mu_{30}^{-8/7}.
\label{eqn:14}
\end{equation}
The above equation relates the IC efficiency with $\dot{M}_{\rm{acc}}$. With a typical assumption that the IC efficiency is $10\%$, $M_{\rm{c}}=1.5\,M_\odot$, $\mu_{30}=1$, $\Gamma_i=10^3$, $\alpha=0.1$, the $\dot{M}_{\rm{acc}}\approx3.5\times10^{15}$\,g/s. Note that $\dot{M}_{\rm{acc}}$ is calculated under the isotropic soft photon field assumption. When the anisotropic soft photon field is considered (anisotropic IC, AIC), the needed $\dot{M}_{\rm{acc}}$ will be higher. The estimation here will serve as a insight of the spectrum calculation in section 4.
\begin{table*}
\centering
\caption{parameters of the model}\label{tab:para}
\label{tab:para}
\begin{tabular}{|c|c|c|c|c|c|c|c|c|c|c|c|}
\hline
 &$K$  & $e$  & $P$       & $i_{\rm{o}}$ & $M_\star$                    & $M_{\rm{c}}$  &$a$                 & $\phi_{\rm{cd},+}$ & $\Delta\phi$\\ 
\hline
1&16.8\,km/s & 0.88 & 315\,days & 80$^\circ$   & 13.2\,$M_\odot$ & 1.5\,$M_\odot$ & $3.29\times10^8$\,km & 170$^\circ$        & 2.2$^\circ$ \\
\hline
2&19.2 & 0.86 & 315 & 70$^\circ$& 13.2 &2.0 & $3.32\times10^8$ & 170$^\circ$ & 2.2$^\circ$ \\
\hline
3&22.5 & 0.88 & 315 & 75$^\circ$& 16.1 &2.4 & $3.55\times10^8$ & 170$^\circ$ & 2.2$^\circ$ \\
\hline
\end{tabular}
\end{table*}
\section{Mass transfer and accretion rate}
The circumstellar material (CM) transfers to the compact star via Bondi-Hoyle process \citep{1944MNRAS.104..273B}. The transfer rate is:
\begin{equation}
\dot{M}_{\rm{transf}}=\eta\pi r_{\rm{BH}}^2\rho v_{\rm{rel}},
\end{equation}
where $\eta$ accounts for the inefficiency of the Bondi-Hoyle accretion, $v_{\rm{rel}}$ is the relative velocity between the neutron star and the circumstellar material, $\rho$ is the density of CM and
\begin{equation}
r_{\rm{BH}}=\frac{2GM_{\rm{c}}}{v_{\rm{rel}}^2}
\end{equation}
is the radius of Bondi-Hoyle accretion when $v_{\rm{rel}}$ is much larger than the sound speed of CM. $v_{\rm{rel}}$ depends on the orientation and inclination of the CD. 

However the collision between the pulsar wind and the stellar wind will alter $v_{\rm{rel}}$ at a shock front radius:
\begin{equation}
r_{\rm{s}}=\sqrt{\frac{L_{\rm{pw}}}{4\pi\rho v^2_{\rm{rel}}c}}.
\end{equation}

Therefore, a necessary condition for accretion is $r_{\rm{BH}}/r_{\rm{s}}>1$:
\begin{equation}
\dot{M}_{\rm{transf}}=
\begin{cases}
0 & r_{\rm{BH}}<r_{\rm{s}}\\
\pi r_{\rm{BH}}^2\rho v_{\rm{rel}} & r_{\rm{BH}}\ge r_{\rm{s}}
\end{cases}\label{eqn:tranfer}
\end{equation}
When the neutron star is outside the CD, evaluation from equation (\ref{eqn:wdv}, \ref{eqn:wdd}) gives $r_{\rm{BH}}/r_{\rm{s}}\ll1$. As a result, mass transfer is impossible outside the CD.

\begin{figure}
\centering
\subfigure{
\includegraphics[width=8cm]{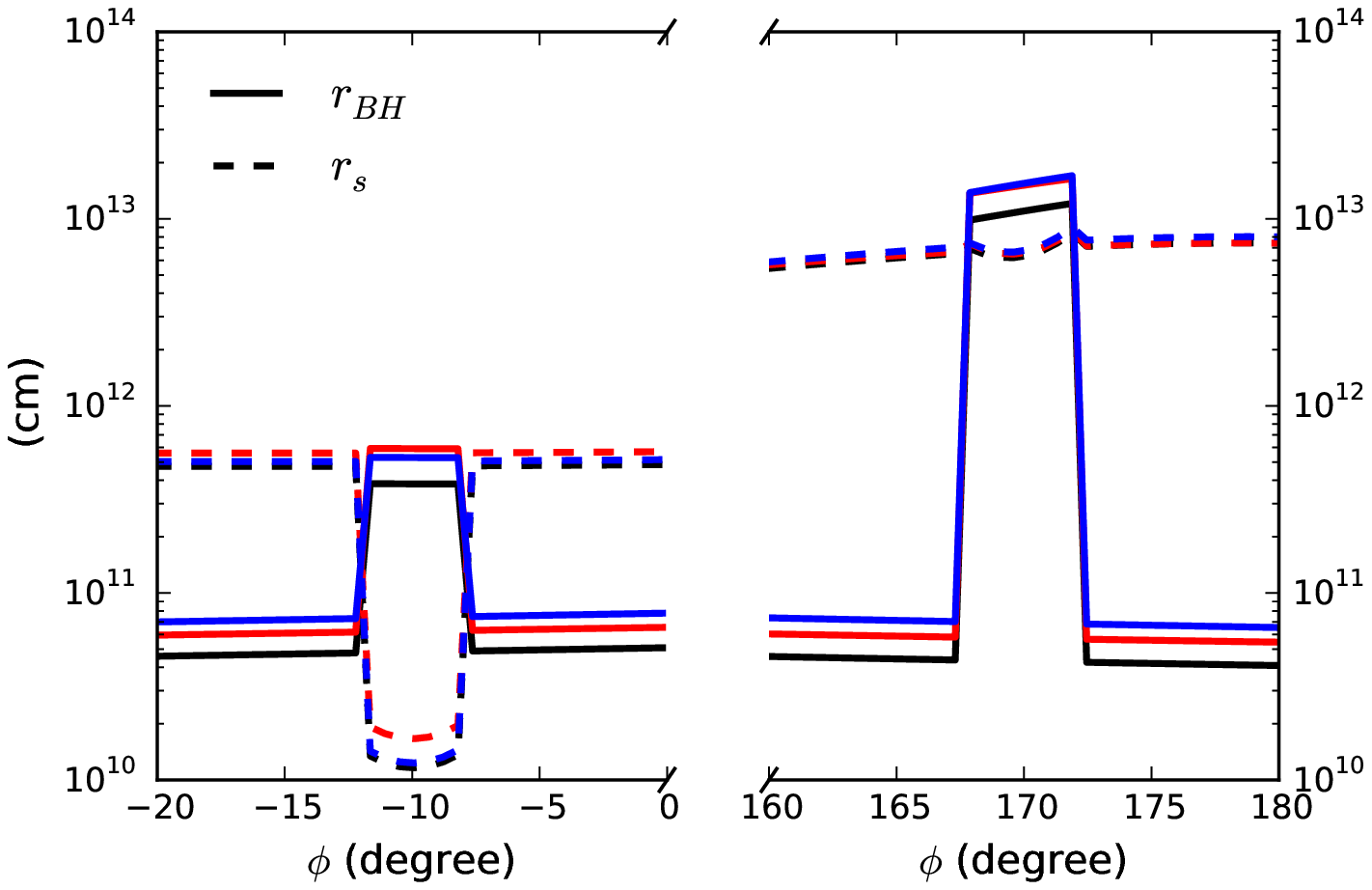}}\hfill
\subfigure{\adjustbox{trim={0.01\width} {.3\height} {0.34\width} {.3\height},clip}{
\includegraphics[width=8cm]{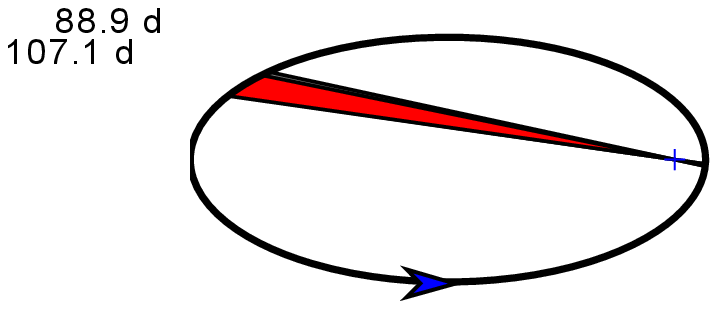}}}
\caption{\textbf{Upper panel:} The $r_{\rm{s}}$ (dashed) and $r_{\rm{BH}}$ (solid) as function of $\phi$ for three sets of parameters, where $\phi$ is the true anomaly. Black lines are for parameters set 1, red lines are for set 2 and blue lines are for set 3. \textbf{Bottom panel:} The illustration of the orbit and the position of the disc (grey shade region). The red shades are regions where mass transfer takes place. The number beside the orbit is the days after periastron of the start and end of the mass transfer. The sense of orbit is counter-clockwise, as indicated with the arrow. Both panels are plotted with $n=3.5$ and $\rho_0=1\times10^{-9}$\,g/cm$^3$}\label{fig:double}
\end{figure}

We have assumed in above section that $\xi=10\%$, and \cite{Li2017} found that the HE luminosity was $\sim2.7\times10^{33}$\,ergs/s. As a result, we suppose the pulsar wind luminosity$L_{\rm{pw}}=2.7\times10^{34}$\,ergs/s.
In figure (\ref{fig:double}) we plot $r_{\rm{s}}$ and $r_{\rm{BH}}$ as functions of the true anomaly (the upper panel). We also visualize the orbit and the location of the CD in the bottom panel (see details in the caption of the figure).

The transferred materials need to have enough angular momenta for an accretion disc to form. When accrete from the CD, the specific angular momenta of the transferred mass originate from the gradients of the velocity and the density:
\begin{equation}
l=\frac{(GM_{\rm{c}})^2}{v^3_{\rm{rel}}}\big(\frac{|\nabla v_{\rm{rel}}|}{v_{\rm{rel}}}+\frac{|\nabla \rho_{\rm{cd}}|}{\rho_{\rm{cd}}}\big),
\label{eqn:l}
\end{equation}
We evaluate the circular radius with $R_{\rm{circ}}=l^2/(GM_{\rm{c}})$ and equation (\ref{eqn:l}), and find that $R_{\rm{circ}}$ is orders of magnitudes larger than the typical light cylinder radius. Therefore the angular momenta are enough.

\begin{figure*}
\includegraphics[width=0.6\textwidth]{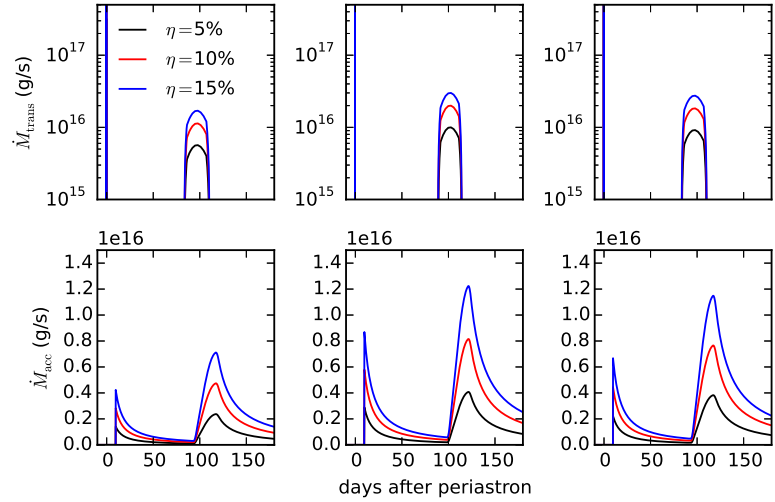}
\caption{\textbf{Upper panels:} Mass transfer rate; \textbf{Lower panels:} The accretion rate corresponding to the mass transfer rate in the upper panel. In each panel, different $\eta$ are indicated with different color. The left, middle and right columns correspond to parameter set 1, 2 and 3 respectively.}\label{fig:three}
\end{figure*}

If the duration of mass transfer is much larger than the viscosity time scale, we expect an stable accretion disc establishing with the accretion rate equal to the mass transfer rate. On the other hand, if the duration of accreting is much less than the viscosity time scale, we expect the accretion disc is transient and the accretion rate at the inner edge of the accretion disc is described as follows \cite{1988Natur.333..523R}:
\begin{equation}
\dot{M}_{\rm{acc},\delta}\propto
\begin{cases}
0 & t<\tau\\
(\frac{t}{\tau})^\beta & t\ge\tau
\end{cases}\label{eqn:resp}
\end{equation}
where $\tau\sim10$\,days is the viscosity time scale and $\beta=-19/16$ for an electron scattering dominated disc opacity \citep{1990ApJ...351...38C}. In the case of this system under the chosen parameters, the duration is about the same order of magnitude with the viscosity time scale. Equation \ref{eqn:resp} can serve as the response of a impulse mass transfer, and the accretion rate corresponding to arbitrary mass transfer rate as function of time $\dot{M}_{\rm{trans}}(t)$ is the convolution between $\dot{M}_{\rm{trans}}(t)$ and equation \ref{eqn:resp}:
\begin{equation}
\dot{M}_{\rm{acc}}(t)=\mathcal{C}on(\dot{M}_{\rm{trans}},\dot{M}_{\rm{acc},\delta}).
\end{equation}
We plot the $M_{\rm{trans}}$ and $M_{\rm{acc}}$ as functions of time, under different $\eta$ in figure \ref{fig:three}.

We expect that the accreted matter does not reach the neutron star, or at least not for a long time. It is because that matter accreted onto the neutron star will quench the pulsar wind, thus destroy the emission mechanism. The accreted matter is expected to be stopped by the propeller effect of the magnetic field. The condition for the propeller effect is that the fastness parameter $f>1$, which is defined as $f=\Omega_\star/\Omega_K(r_{\rm{A}})$, where $\Omega_\star$ is the angular velocity of the pulsar, and $\Omega_{\rm{K}}(r_{\rm{A}})$ is the angular velocity of Kepler motion at Alfven radius $r_{\rm{A}}$. With the assumption of a typical magnetic field strength $B=10^{12}$\,G and spin frequency $\nu>1$, the condition of the fastness parameter is satisfied.

\section{The HE spectrum and the light curve}
In this section, we work on the HE spectra from AIC.
With $\dot{M}_{\rm{acc}}$ evaluated in the above way, we calculate the $t_{\rm{in}}$ and $r_{\rm{in}}$ as functions of time, with Equations (\ref{eqn:temper} and \ref{eqn:alfven}). For simplicity, we assume the accretion disc is face on. Since the soft photons are mainly contributed by the inner edge of the accretion disc, we make a further simplification that the soft photons come from a ring with width of $r_{\rm{in}}$ at $r_{\rm{in}}$. Therefore:
\begin{equation}
n_{\rm{ph}}(\epsilon_0,l)=\frac{4\pi l}{h^3c^3}\frac{\epsilon_0^2r_{\rm{in}}^2}{(r^2_{\rm{in}}+l^2)^{3/2}(\exp\frac{\epsilon_0}{kT_{\rm{in}}}-1)}.
\end{equation}

Assuming the velocities of the electrons in the pulsar wind zone $dld\Omega$ follows a Gaussian distribution:
\begin{equation}
\frac{dN(\Gamma)}{d\Gamma dld\Omega}=N_0\exp\big(-\frac{(\Gamma-\Gamma_0)^2}{2\Delta\Gamma^2}\big),
\end{equation}
where $N_0$ is normalized to the number of electrons per unit length around $l$:
\begin{equation}
\int\frac{N(\Gamma)}{d\Gamma d\Omega dl}\Gamma m_{\rm{e}}c^2d\Gamma=\frac{L_{\rm{pw}}}{4\pi c},\label{eqn:normal}
\end{equation}
From equation (\ref{eqn:normal}), we have:
\begin{equation}
N_0=\frac{L_{\rm{pw}}}{4\pi c\sqrt{2\pi}}\frac{1}{\Delta\Gamma\Gamma_0m_{\rm{e}}c^2}.
\end{equation}
In principle, $\Delta\Gamma$ and $\Gamma_0$ can all be function of $l$.
The spectrum of AIC is:
\begin{eqnarray}
\begin{aligned}
&\frac{dN_{\rm{tot}}}{dAdtd\epsilon_1}=\\
&\frac{1}{D^2}\int dl\int d\Omega(l)\int d\Gamma\int d\epsilon_0n_{\rm{ph}}(\epsilon_0,l)\frac{dN(\Gamma)}{d\Gamma dld\Omega}\frac{dN_{\Gamma,\epsilon_0}}{dtd\epsilon_1}.
\label{eqn:spectrum}
\end{aligned}
\end{eqnarray}
where $dN_{\Gamma,\epsilon_0}/dtd\epsilon_1$ is the kernel of AIC in the Thomson region (which is given in equation (7) of \citealt{2008A&A...477..691D}), $\epsilon_0$ is the energy of the soft photon, and $\epsilon_1$ is the energy  of up-scattered photon. $\int d\Omega(l)$ integrates over the all the coming direction of the soft photons at wind zone $l$.
The coming angle of the soft photons are concentrated around:
\begin{equation}
\cos\theta_0\approx-\frac{l}{\sqrt{r_{\rm{in}}^2+l^2}},
\end{equation}
and
\begin{equation}
\int d\Omega\approx2\pi\frac{l}{\sqrt{r^2_{\rm{in}}+l^2}}\frac{r^2_{\rm{in}}}{r^2_{\rm{in}}+l^2}.
\end{equation}

Since the AIC is most efficient in a small distance range around the inner edge of the accretion disc, we assume $\Gamma_0$ and $\Delta\Gamma$ to be constant.

\begin{figure}
\centering
\includegraphics[width=7cm]{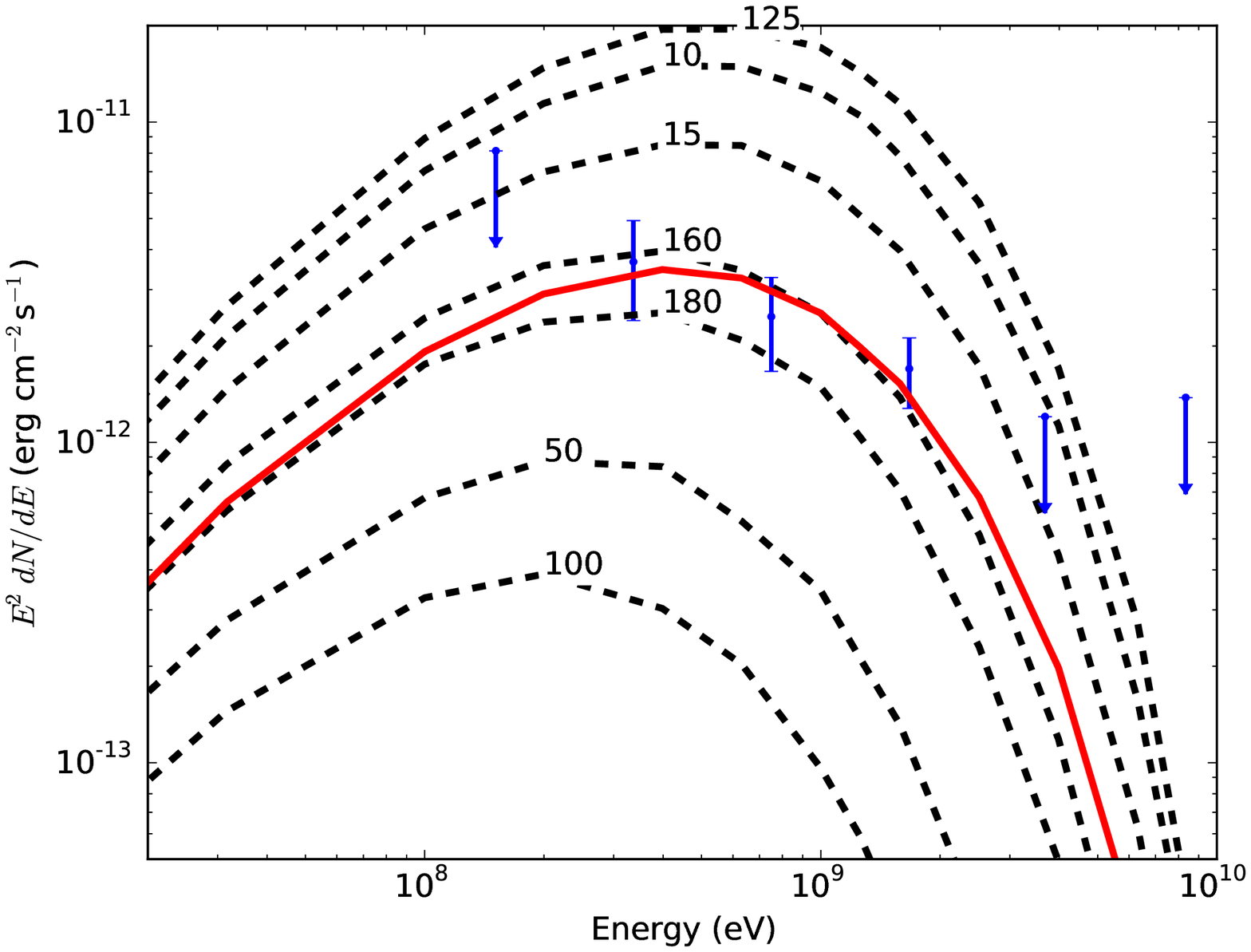}
\caption{\textbf{The evolution of the SEDs of AIC:} black dashed curves are SED at different times. The numbers labeled besides each curve indicates the corresponding days after the periastron. The SEDs of AIC are calculated with parameter set 2 in Table \ref{tab:para}, \etavalue, \gammavalue and \deltagammavalue. The blue dots with error bar are observed data of \protect\cite{Li2017}, and those with downward arrows are observed upper limits. The red solid curve is the phase averaged SED.}
\label{fig:evospec}
\end{figure}

\begin{figure*}
\includegraphics[width=0.8\textwidth]{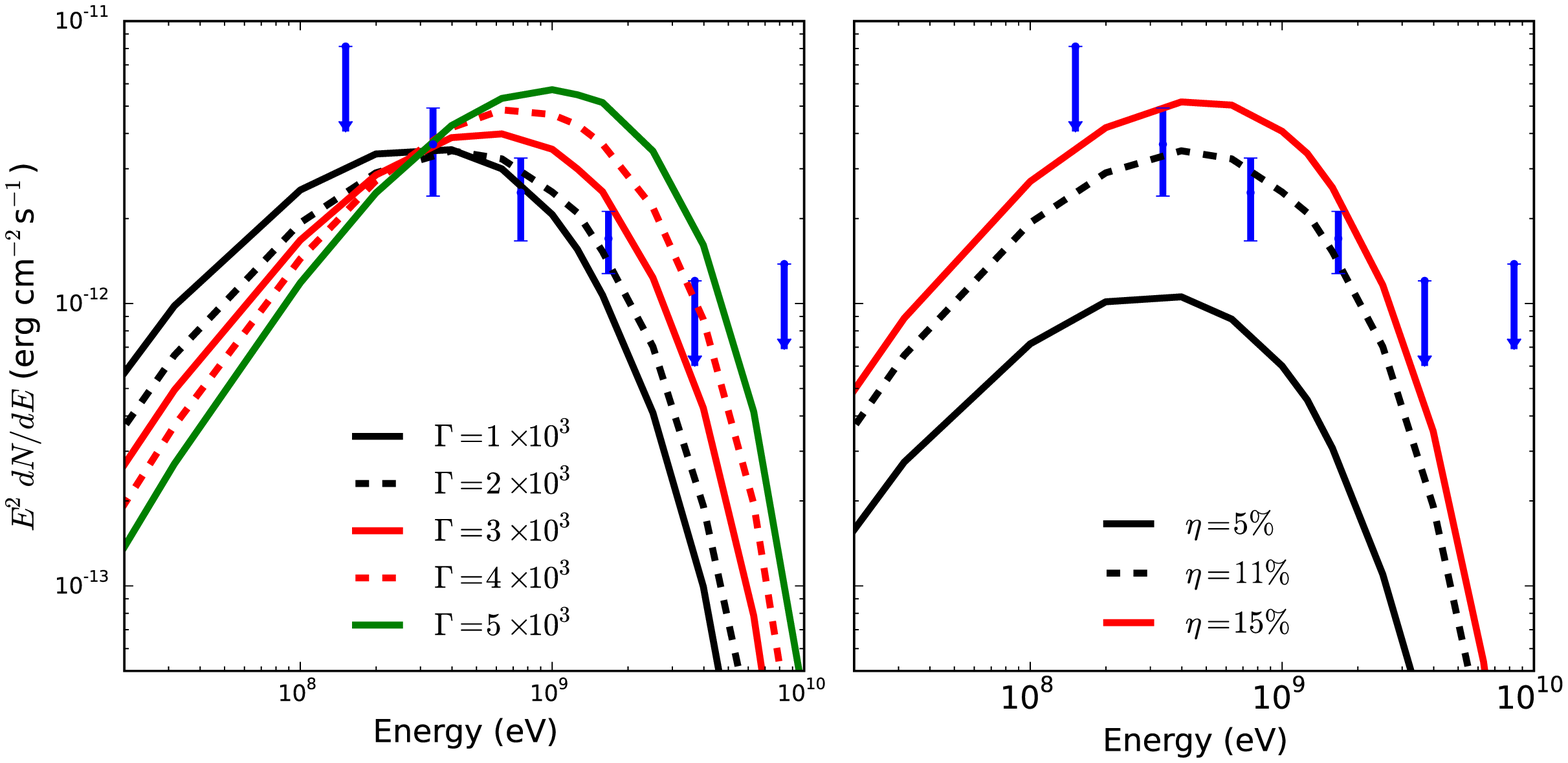}
\caption{\textbf{Same figures as figure \ref{fig:evospec}, but with different $\Gamma_0$ and $\eta$}. The SEDs in the left panel correspond to different $\Gamma_0$ (in different line colors and styles as shown in the legend) with $\eta=11\%$. The SED in the right panel correspond to different $\eta$ with $\Gamma_0=2\times10^3$.}\label{fig:replace}
\end{figure*}

The evolution of the SEDs of AIC are plotted in figure (\ref{fig:evospec}) as black dashed curves with the dates labeled beside each curve. These SEDs are calculated with parameter set 2 in Table \ref{tab:para}, \etavalue, \gammavalue and \deltagammavalue. The phase-averaged SED is plotted as the red solid line. For comparison, we plot in figure \ref{fig:evospec} the observed SED from \cite{Li2017}. The observed SED is orbital phase averaged. We plot in figure \ref{fig:replace} the SEDs with different choices of $\Gamma_0$ and $\eta$.

\begin{figure}
\centering
\includegraphics[width=7cm]{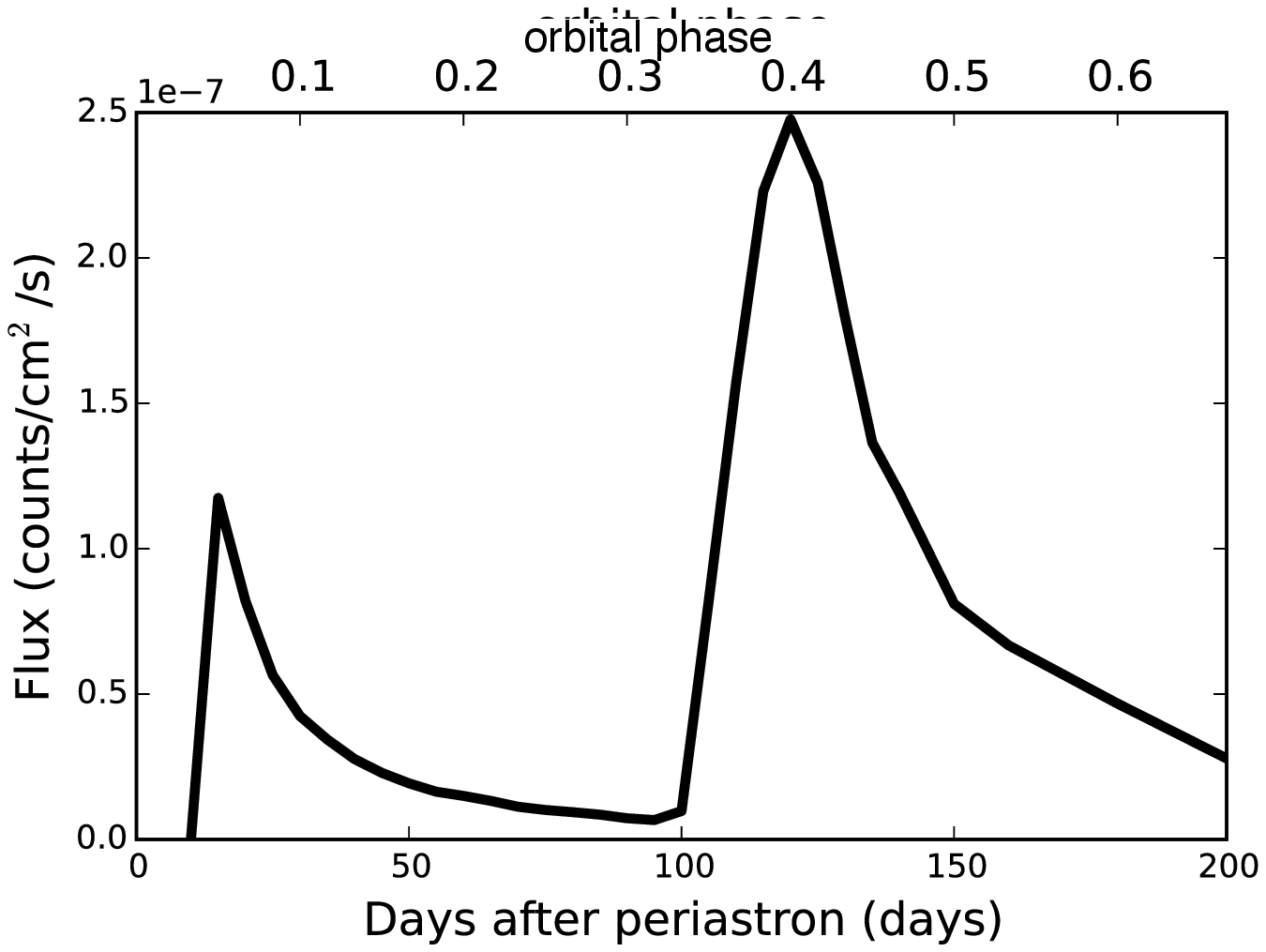}
\caption{The light curve in the HE range. The parameters are the same as in figure \ref{fig:evospec}.}\label{fig:lc}
\end{figure}

Integrating the spectra over the HE range gives the light curve. The light curve is plotted in figure \ref{fig:lc}.
\section{Summary and Conclusions}
In this paper, we study the possibility that the compact companion of HESS J0632+057 is a pulsar. Under the current observational constraints, the probabilities that the compact companion has a mass less than $1.5\,M_\odot$, $2.0\,M_\odot$ and $2.4\,M_\odot$ are $\sim8\%$, $\sim17\%$ and $\sim28\%$ respectively.

In the pulsar framework, we use a new mechanism that could account for the orbital modulated GeV emission from HESS J0632+057  (cf. \citealt{2017ApJ...836..241T,Yi2017}). The pulsar companion travels through the circumstellar disc of the main sequence star twice in each orbit, when some of the matter is gravity-captured. The captured matter develops an accretion disc around the pulsar, and the soft photons from which are inverse Compton scattered by the pulsar wind as the GeV emission from the system.

Under assumptions of parameters within observational constraints, the above-mentioned mechanism can produce SED and light curve in accordance with current observations (see figures \ref{fig:evospec} and \ref{fig:lc}).

One major prediction of the proposed mechanism is that the light curve has two peaks, the larger one is at around 0.4 after periastron, while the smaller one is between phases 0 and 0.1, with integrated flux one forth of the larger one.
\section*{Acknowledgement}
This work is partially supported by a GRF grant under 17302315.

\end{document}